\documentclass[twocolumn,letter]{jpsj3}
\usepackage{graphicx}
\usepackage{bm}
\usepackage{cite}
\usepackage{color}
\usepackage{amssymb}
\usepackage{amsmath}
\usepackage{revsymb}

\title{Reversed Crystal-Field Splitting and Spin-Orbital Ordering in $\alpha$-Sr$_2$CrO$_4$}

\author{
Takashi Ishikawa$^1$, 
Tatsuya Toriyama$^1$, 
Takehisa Konishi$^2$, 
Hiroya Sakurai$^3$, and 
Yukinori Ohta$^1$\thanks{ohta@faculty.chiba-u.jp}
}
\inst{
$^1$Department of Physics, Chiba University, Chiba 263-8522, Japan \\
$^2$Graduate School of Advanced Integration Science, Chiba University, Chiba 263-8522, Japan \\
$^3$National Institute for Materials Science, Tsukuba, Ibaraki 305-0044, Japan
}

\abst{
The origin of successive phase transitions observed in the layered perovskite $\alpha$-Sr$_2$CrO$_4$ 
is studied by the density-functional-theory-based electronic structure calculation and mean-field 
analysis of the proposed low-energy effective model.  We find that, despite the fact that the CrO$_6$ 
octahedron is elongated along the $c$-axis of the crystal structure, the crystal-field level 
of nondegenerate $3d_{xy}$ orbitals of the Cr ion is lower in energy than that of doubly degenerate 
$3d_{yz}$ and $3d_{xz}$ orbitals, giving rise to the orbital degrees of freedom in the system with a 
$3d^2$ electron configuration.  We show that the higher (lower) temperature phase transition is 
caused by the ordering of the orbital (spin) degrees of freedom.  
}

\begin{document}
\maketitle

%%%%%%%%%%%%%%%%%%%%%
%\section{Introduction}
%%%%%%%%%%%%%%%%%%%%%

The orbital degrees of freedom in transition-metal compounds have long been one of the 
major themes in the physics of strongly correlated electron systems \cite{khomskii,tokura,imada}.  
The simplest example is the cubic perovskite structure, where the transition-metal ion 
surrounded by six ligand ions forms an octahedron and the corresponding crystal field splits 
the energy levels of the five $d$ orbitals into triply degenerate $t_{2g}$ orbitals and doubly 
degenerate $e_g$ orbitals.  Moreover, in a layered perovskite with the K$_2$NiF$_4$-type 
crystal structure, the octahedron is elongated along the $c$-axis and the triply degenerate 
$t_{2g}$ levels further split into low-energy doubly degenerate $d_{xz}$ and $d_{yz}$ orbitals 
and the high-energy nondegenerate $d_{xy}$ orbital \cite{fazekas}.  No one would have doubted 
this simple law of the crystal field theory.  

In this paper, we show that a serious deviation from this simple law occurs in the Mott insulator 
$\alpha$-Sr$_2$CrO$_4$.  This material has the K$_2$NiF$_4$-type crystal structure with 
CrO$_6$ octahedra elongated along the $c$-axis and with a $3d^2$ electron configuration 
\cite{kafaras,baikie}.  
Therefore, one would naturally expect that two electrons occupy the lowest doubly degenerate 
$t_{2g}$ orbitals forming an $S=1$ spin due to Hund's rule coupling, so that only 
the antiferromagnetic N\'eel ordering of $S=1$ spins occurs at the N\'eel temperature 
$T_{\rm N}$, without any orbital ordering \cite{baikie}.  
Surprisingly, however, a recent experimental study \cite{sakurai} revealed that two phase 
transitions occur successively at 112 and 140 K, releasing nearly the same amount of 
entropy.  
The lower-temperature phase transition was ascribed to N\'eel ordering by magnetic 
measurement, but the cause of the higher-temperature one (denoted as $T_{\rm S}$) 
remains a mystery from the experiment \cite{sakurai,sugiyama}.  

In what follows, using the density-functional-theory (DFT)-based electronic structure 
calculations, we will show that, in $\alpha$-Sr$_2$CrO$_4$, the crystal-field level of 
nondegenerate $3d_{xy}$ orbitals of the Cr ion is in fact lower in energy than that of doubly 
degenerate $3d_{yz}$ and $3d_{xz}$ orbitals.  Therefore, in this system with the $3d^2$ electron 
configuration, the orbital degrees of freedom are indeed active.  More precisely, this system 
can be modeled as the Kugel-Khomskii-type spin-orbital subsystem \cite{kugel1,kugel2} 
consisting of the $d_{xz}$ and $d_{yz}$ orbitals of the Cr ion, which couples to the antiferromagnetic 
Heisenberg subsystem consisting of the $d_{xy}$ orbitals, via the on-site Hund's ferromagnetic 
exchange interaction.  This description is also justified by the DFT calculation allowing 
for spin polarization, which predicts the spin- and orbital-ordered ground state in this material.  
%
%We discuss the origin of the reversal of the crystal-field levels in terms of the imbalance 
%between the atomic energy levels of the in-plane and apical oxygens and the distortion of 
%the CrO$_6$ octahedra.  
%
The mean-field analysis of the proposed low-energy effective model, taking into account 
the Jahn-Teller distortion, can explain the successive phase transitions observed in 
$\alpha$-Sr$_2$CrO$_4$; i.e., the higher (lower) temperature phase transition is caused by 
the ordering of the orbital (spin) degrees of freedom.  
To the best of our knowledge, this is one of the rare examples \cite{spinel} of the reversal of 
the crystal-field levels in transition-metal compounds, which is proved by the solid experimental 
consequence.  

%%%%%%%%%%%%%%%%%%%%%%%%%%%%%
%\section{Computational Details}
%%%%%%%%%%%%%%%%%%%%%%%%%%%%%

 We employ the WIEN2k code \cite{wien2k} based on the full-potential linearized 
augmented-plane-wave method for our DFT calculations.  We present calculated results 
obtained in the generalized gradient approximation (GGA) for electron correlations with 
the exchange-correlation potential of Ref.~13.  To improve the description 
of electron correlations in the Cr $3d$ orbitals, we use the rotationally invariant version 
of the GGA+$U$ method with the double-counting correction in the fully localized limit 
\cite{anisimov,liechtenstein}, where we choose $U$ to reproduce the band gap observed 
in experiment.  The spin polarization is allowed when necessary.  The spin-orbit interaction 
is not taken into account.  
We use the crystal structure measured at a low temperature \cite{sakurai}, which has the 
tetragonal symmetry (space group $I4/mmm$) with the lattice constants $a=3.821$ and 
$c=12.492$ in units of \AA.  
%The unit cell contains two Sr ions, one Cr ions, and four O ions.  
There are one (two) crystallographically inequivalent Cr (O) ions.  
Because no significant structural changes were detected at the phase-transition points 
\cite{sakurai}, we assume the same crystal structure for both high- and low-temperature 
phases, but the unit cell is extended to a $\sqrt{2}\times\sqrt{2}\times 1$ supercell, 
allowing for antiferromagnetic spin polarization in the low-temperature phase.  
In the self-consistent calculations, we use 14850 ${\bm k}$-points in the irreducible part of 
the Brillouin zone.  Muffin-tin radii ($R_{\rm MT}$) of 2.35  (Sr), 1.88 (Cr), and 1.70 (O) Bohr 
are used and a plane-wave cutoff of $K_{\rm max}=7.00/R_{\rm MT}$ is assumed.  
%We use the codes VESTA \cite{momma} and XCrySDen \cite{kokalj} for graphical purposes.  

%%%%%%%%%%%%%%%%%%%%%%%%%%%%%
%\section{Results of calculation}
%%%%%%%%%%%%%%%%%%%%%%%%%%%%%

\begin{figure}[thb]
\begin{center}
\includegraphics[width=7.8cm]{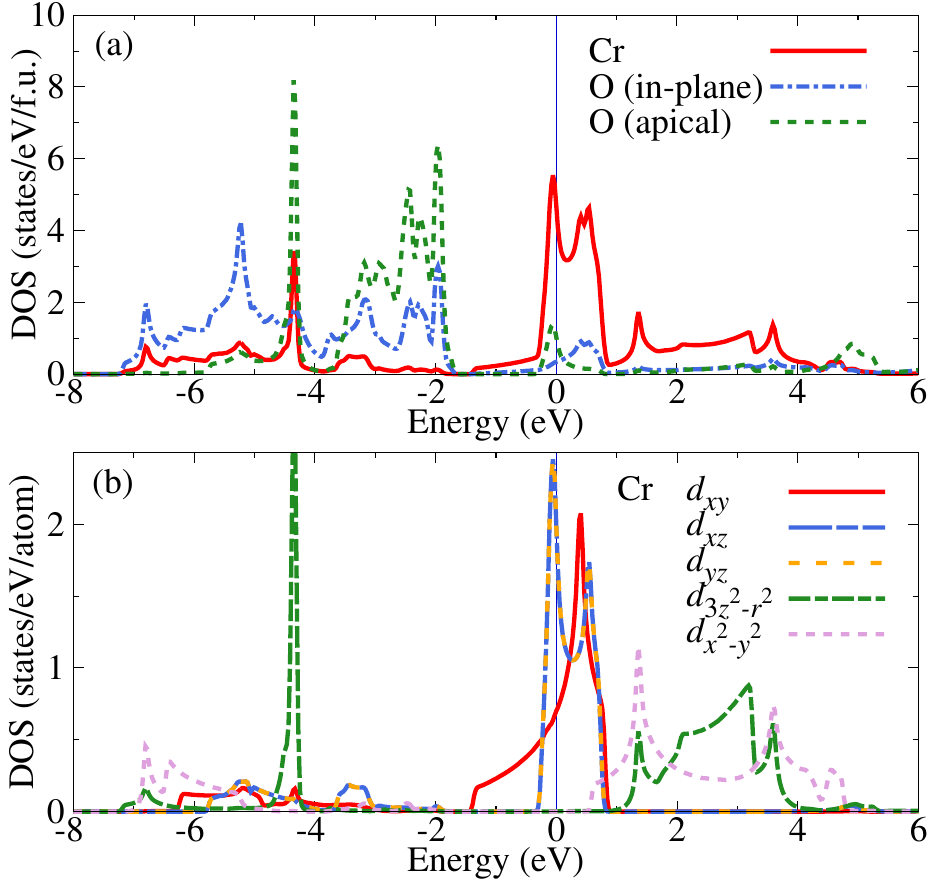}
\caption{(Color online) 
(a) Calculated total and partial DOSs (per formula unit, f.u.) in the 
hypothetical paramagnetic metallic state of $\alpha$-Sr$_2$CrO$_4$.  
(b) Calculated partial DOSs projected onto the five $d$ orbitals of Cr.  
The vertical line in each panel represents the Fermi level.  
}\label{fig1}
\end{center}
\end{figure}

%\subsection{Calculated crystal field levels}

\begin{figure}[thb]
\begin{center}
\includegraphics[width=8.7cm]{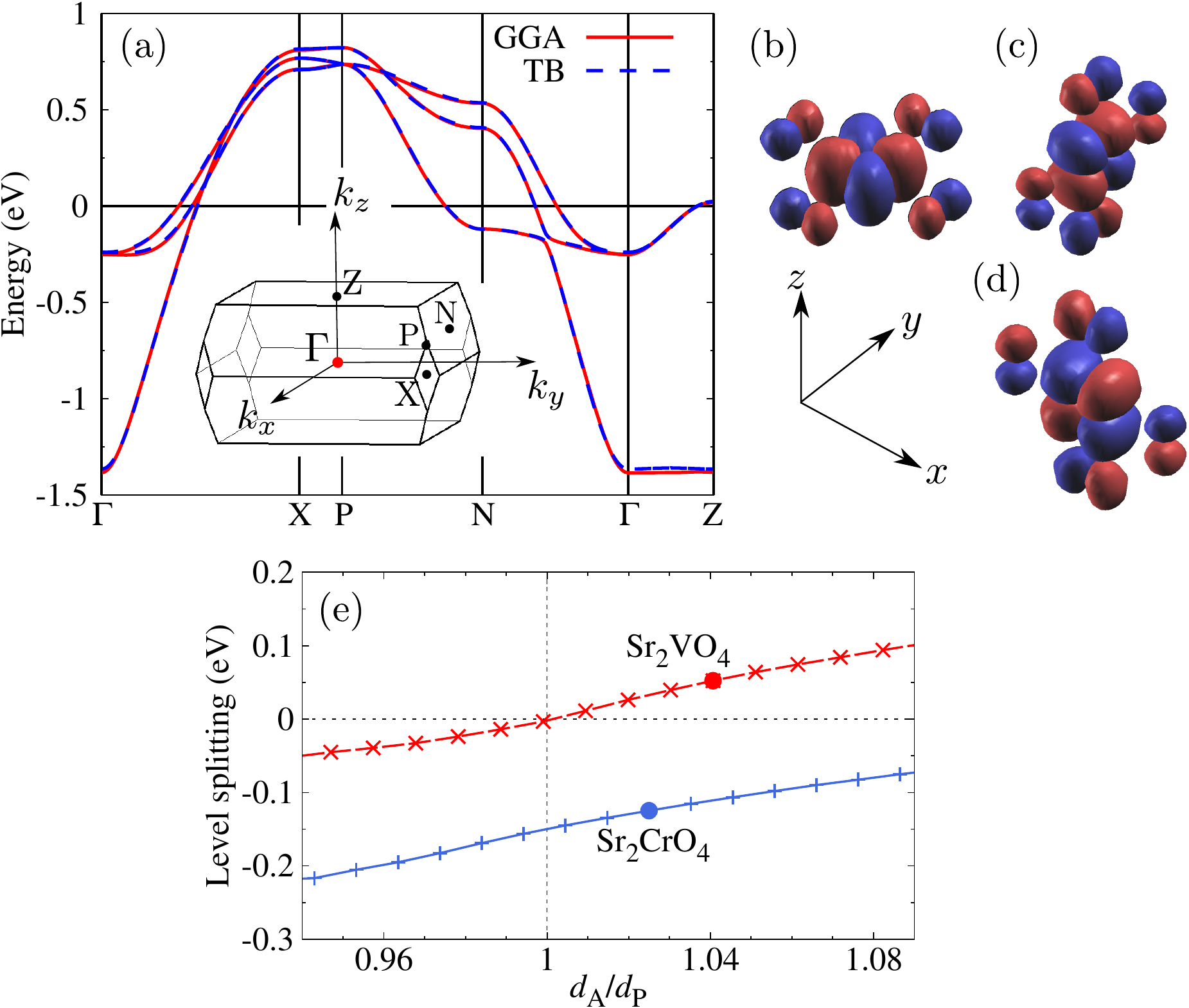}
\caption{(Color online) 
(a) Calculated band dispersions near the Fermi level compared with the tight-binding 
bands of the maximally localized Wannier orbitals.  The horizontal line indicates the Fermi 
level.  The inset shows the Brillouin zone.  
(b) Illustration of the maximally localized Wannier functions for the $3d_{xy}$ orbital 
of Cr admixed with the $2p_x$ and $2p_y$ orbitals of O(P).  
(c) Same as in (b) but for the $3d_{yz}$ orbital admixed with $2p_z$ of O(P) and 
$2p_y$ of O(A).  
(d) Same as in (c) but for the $3d_{xz}$ orbital.  
(e) Calculated energy-level splitting $\varepsilon(d_{xy})-\varepsilon(d_{yz})$ of the 
maximally localized Wannier orbitals as a function of $d_{\rm A}/d_{\rm P}$.  
The results for Sr$_2$VO$_4$ are also shown for comparison.  
The dots indicate the results obtained for the experimental bond lengths.  
}\label{fig2}
\end{center}
\end{figure}

First, let us discuss the hypothetical paramagnetic metallic phase of $\alpha$-Sr$_2$CrO$_4$ 
obtained under the assumption of no spin polarization.  The calculated results for the density 
of states (DOS) are shown in Fig.~\ref{fig1}, where we find that the states near the Fermi level 
consist mainly of the $3d$ $t_{2g}$ orbitals of the Cr ion.  The orbital-decomposed partial DOSs 
shown in Fig.~\ref{fig1}(b) indicate that the $3d_{xy}$ orbitals form the skew DOS corresponding 
to the tight-binding bands of the two-dimensional square lattice and the $3d_{xz}$ and $3d_{yz}$ 
orbitals form the skew DOSs corresponding to those of the one-dimensional chain.  
The dispersions of the three bands near the Fermi level are shown in Fig.~\ref{fig2}(a), which 
are fitted very well by the tight-binding bands of the three molecular orbitals obtained as the 
maximally localized Wannier functions \cite{mostofi,kunes}.  
%, where we use the WANNIER90 \cite{mostofi} and WIEN2WANNIER \cite{kunes} codes.  
The shapes of the obtained Wannier functions are illustrated in Fig.~\ref{fig2}(b), where we 
find that the $d_{xy}$ molecular orbital is composed of the Cr $3d_{xy}$ atomic orbital with a 
strong admixture of the $2p_\pi$ atomic orbitals of the in-plane oxygens O(P).  In Fig.~\ref{fig2}(c), 
we also show the $d_{yz}$ molecular orbital composed of the Cr $3d_{yz}$ atomic orbital with 
a strong admixture of the $2p_z$ atomic orbitals of O(P) and $2p_y$ atomic orbitals of apical 
oxygen O(A).  

From the above tight-binding fitting, we derive the crystal-field levels of the $t_{2g}$ orbitals.  
The obtained energy-level splitting between the $d_{xy}$ molecular orbital and the $d_{xz}$ and 
$d_{yz}$ molecular orbitals is shown in Fig.~\ref{fig2}(e) as a function of the distance $d_{\rm A}$ 
between the Cr and O(A) ions divided by the distance $d_{\rm P}$ between the Cr and O(P) ions.  
The splitting is calculated by shifting only the positions of O(A) ions from the experimental 
position.  The results of $\alpha$-Sr$_2$VO$_4$ are also shown for comparison.  
We find here that indeed the crystal-field levels are reversed in $\alpha$-Sr$_2$CrO$_4$, 
regardless of the elongation of the CrO$_6$ octahedron examined, which is in sharp contrast 
to the case of $\alpha$-Sr$_2$VO$_4$ \cite{imai,zhou,sugiyama2}.  
We also examined the electronic state of LaSrVO$_4$, which has the same crystal structure 
with the $3d^2$ electron configuration \cite{greedan,longo}; only the antiferromagnetic phase 
transition was observed experimentally \cite{sugiyama,dun}.  
%although the existence of ``orbital fluctuations'' was reported \cite{dun}.  
Our DFT calculation shows that the reversal of the crystal-field levels does not occur in 
this material.  

\begin{figure}[thb]
\begin{center}
\includegraphics[width=8.6cm]{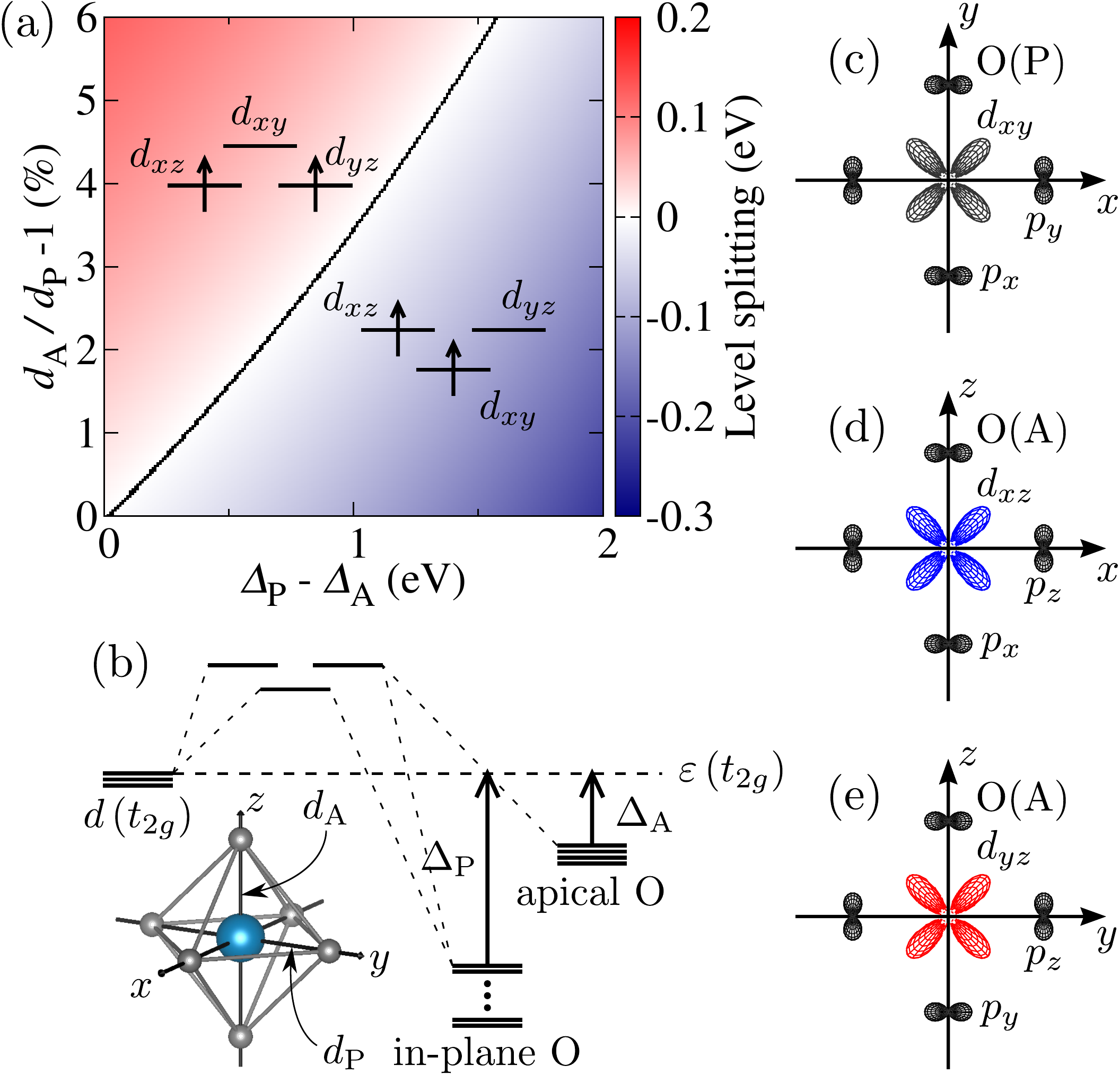}
\caption{(Color online) 
(a) Calculated energy-level splitting in the CrO$_6$ cluster as a function of the imbalance 
between the atomic levels of apical and in-plane O and the distortion of the CrO$_6$ 
octahedron.  
(b) Schematic representation of the energy levels of the atomic and molecular orbitals of 
CrO$_6$ octahedron.  
(c) Schematic representation of the $3d_{xy}$ orbital of Cr and the $2p_x$ and $2p_y$ orbitals 
of O, which form the $d_{xy}$ molecular orbital.  
(d) Same as in (c) but for the $d_{xz}$ molecular orbital.  
(e) Same as in (c) but for the $d_{yz}$ molecular orbital.  
}\label{fig3}
\end{center}
\end{figure}

%\subsection{Origin of the reversed crystal-field splitting}

The origin of this reversal of the crystal-field levels manifests itself in the energy levels 
of the CrO$_6$ cluster.  Here, we take into account the three $t_{2g}$ atomic orbitals of Cr 
and twelve $2p$ atomic orbitals of O(P) and O(A), which are linked by $\pi$ bonds.  
The $e_g$ atomic orbitals and $2p$ atomic orbitals linked by $\sigma$ bonds, which 
are orthogonal to the $t_{2g}$ orbitals, are neglected.  The energy levels of these atomic 
orbitals, together with the energy levels of molecular orbitals as the antibonding orbitals 
of the $t_{2g}$ and $2p$ atomic orbitals, are defined in Fig.~\ref{fig3}(b); in particular, we 
define the level differences $\Delta_{\rm A}=\varepsilon(t_{2g})-\varepsilon({\rm O(A)})$ 
and $\Delta_{\rm P}=\varepsilon(t_{2g})-\varepsilon({\rm O(P)})$.  
The calculated results for the crystal-field splitting are shown in Fig.~\ref{fig3}(a), where 
we find that the imbalance between the atomic levels of O(A) and O(P), as well as the 
distortion of the CrO$_6$ octahedron, plays an essential role.  
The energy level of the atomic $2p$ orbitals of O(A) shifts upward and approaches the energy 
levels of the $t_{2g}$ atomic orbital, just as in the negative charge-transfer-gap situation 
\cite{zaanen,khomskii2} occurring in, e.g., CrO$_2$ \cite{korotin} and K$_2$Cr$_8$O$_{16}$ 
\cite{sakamaki,toriyama}, which causes the reversal of the crystal-field levels.  

%\subsection{Antiferromagetic solution and orbital ordering pattern}

\begin{figure}[thb]
\begin{center}
\includegraphics[width=7.8cm]{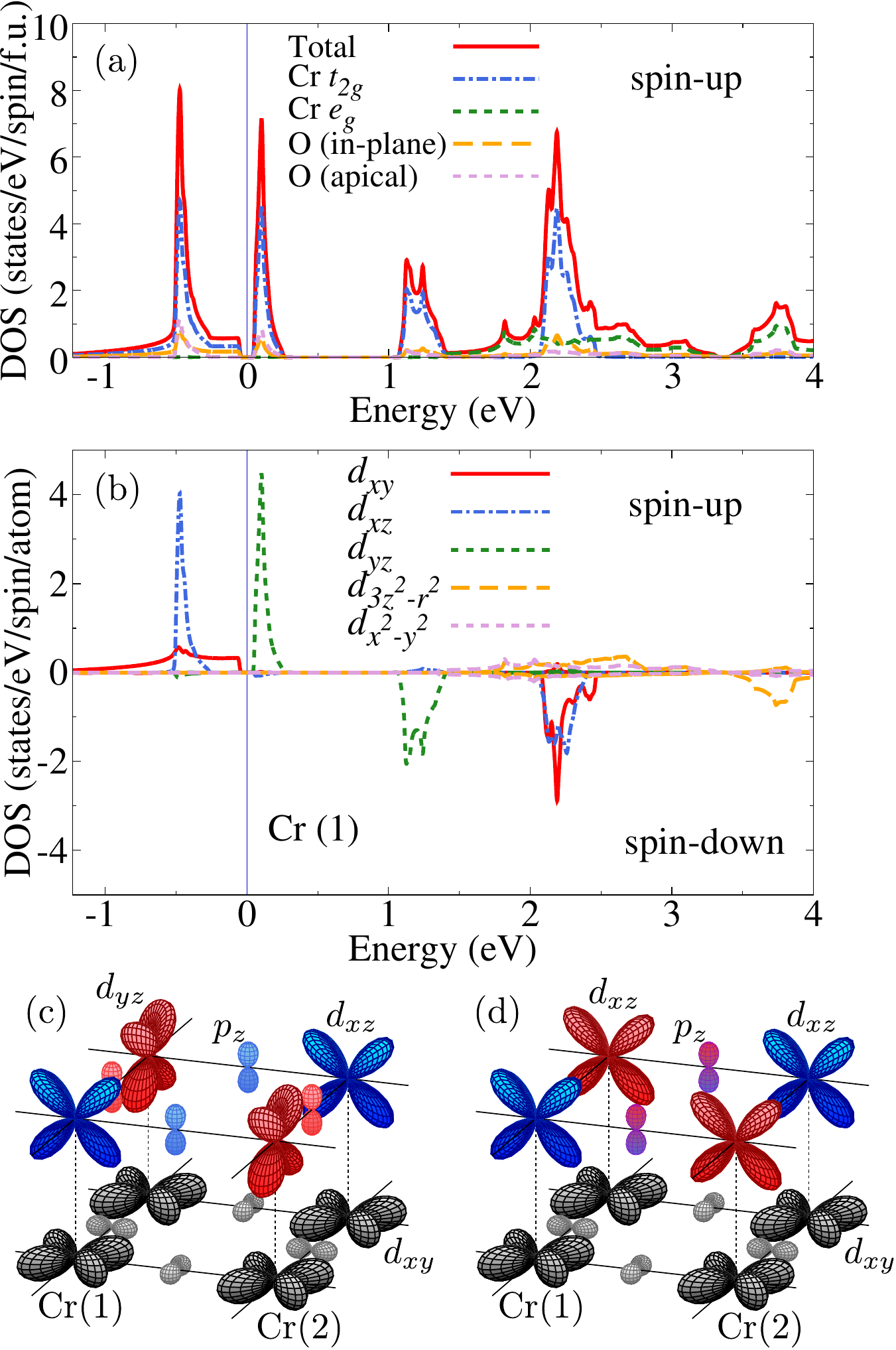}
\caption{(Color online) 
Calculated (a) total and (b) orbital-decomposed partial DOSs in the 
antiferromagnetic phase of $\alpha$-Sr$_2$CrO$_4$.  
The vertical line indicates the Fermi level.  
$U=0.9$ eV is assumed.  
(c) Schematic representation of the ground-state spin-orbital ordering pattern in $\alpha$-Sr$_2$CrO$_4$.  
Note that the $d_{xz}$ and $d_{yz}$ orbitals are illustrated separately from the $d_{xy}$ orbitals.  
(d) Same as in (c) but for the second lowest energy state.  
}\label{fig4}
\end{center}
\end{figure}

Next, let us discuss the symmetry-broken ground state of $\alpha$-Sr$_2$CrO$_4$.  
We carry out the DFT calculation allowing for spin polarization using the GGA+$U$ approach.  
The calculated results are shown in Fig.~\ref{fig4}, where we find that the spin- and 
orbital-ordered ground state is actually realized; i.e., the $3d_{xy}$ orbitals are always occupied 
by one electron, but either the $d_{xz}$ or $d_{yz}$ orbital is occupied by one electron, leading to 
orbital ordering.  This result reinforces the concept of the reversed crystal-field levels 
in $\alpha$-Sr$_2$CrO$_4$ discussed above.  
The calculated spatial pattern indicates that the spin degrees of freedom show the spin 
$S=1$ antiferromagnetic N\'eel ordering as expected, but the orbital degrees of freedom 
also show antiferro ordering, which is not expected from the Kugel-Khomskii theory \cite{kugel1,kugel2}.  
However, we find that the spin-antiferro and orbital-ferro state also appears as a metastable 
state with the second lowest energy; the energy difference is only 0.02122 eV/f.u.  
There might be other low-energy orbital ordering patterns if we use larger supercells; 
the real pattern should eventually be determined experimentally.  

We also find that the energy gap appears in the calculated DOS curves (see Fig.~\ref{fig4}), 
so that the system is insulating in agreement with the experiment where the band gaps of 
0.10 eV (from the resistivity of a bulk sample) \cite{sakurai} and $\sim$0.3 eV (from 
optical measurement on a thin film) \cite{matsuno} were reported.  $U$ can be adjusted 
to reproduce such band gaps without any significant changes in our discussion.  

%\subsection{mean-field analysis}

\begin{figure}[thb]
\begin{center}
\includegraphics[width=8.5cm]{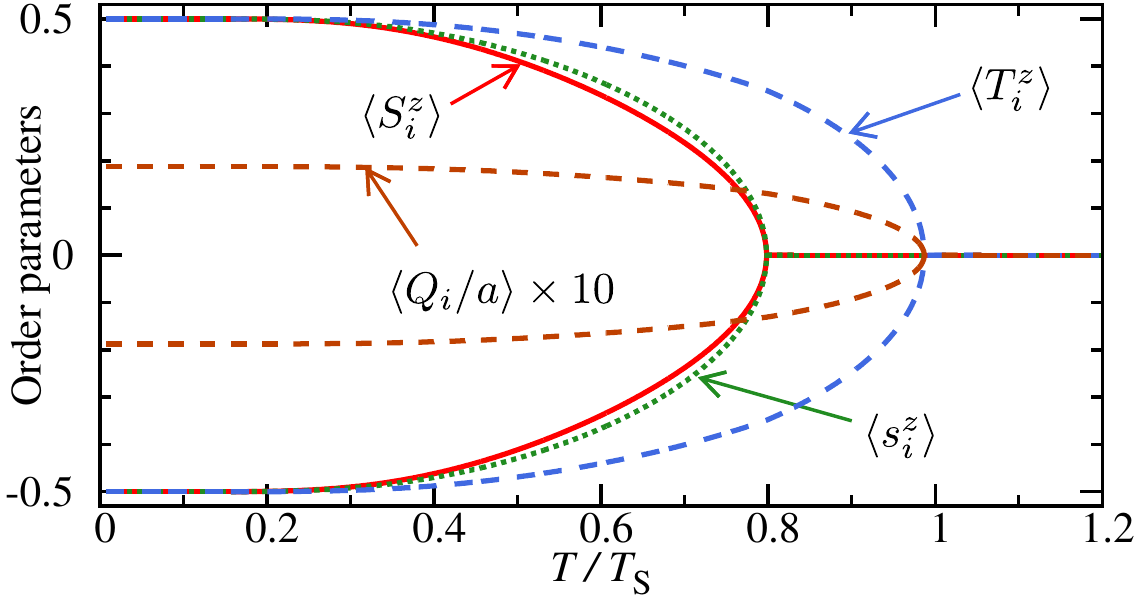}
\caption{(Color online) 
Temperature dependence of the order parameters calculated in the mean-field 
approximation for our effective model.  The parameter values used are 
$K=0.05$, $J_{\rm AF}=0.1$, $J_{\rm H}=0.5$, $ga=15$, $ka^2=2500$, and $\lambda a^2=525$ 
in units of eV, which result in $T_{\rm S}=1390$ K and $T_{\rm N}=1110$ K.  
}\label{fig5}
\end{center}
\end{figure}

Finally, let us discuss the low-energy effective model describing the spin and orbital orderings 
in $\alpha$-Sr$_2$CrO$_4$.  This model represents the two-dimensional CrO$_4$ plane made of 
(i) the Kugel-Khomskii subsystem consisting of the Cr $3d_{xz}$ and $3d_{yz}$ orbitals (hybridized with O $2p$ orbitals), leading to the orbital ordering, 
(ii) the antiferromagnetic Heisenberg subsystem consisting of the $3d_{xy}$ orbital (hybridized with O $2p$ orbitals), leading to the robust antiferromagnetic background, 
(iii) the Hund's ferromagnetic exchange interaction acting between the electron on the $3d_{xy}$ orbital and electron on the $3d_{xz}$ or $3d_{yz}$ orbital in the same Cr site, 
and (iv) the Jahn-Teller coupling with the lattice degrees of freedom.  
This model in the strong-coupling version may be defined by the Hamiltonian 
\begin{align}
{\cal H}_{\rm eff}&=K\sum_{\langle ij\rangle}\Big(2{\bm S}_i\cdot{\bm S}_j+\frac{1}{2}\Big)
\Big(2{\bm T}_i\cdot{\bm T}_j+\frac{1}{2}\Big)\notag\\
&+J_{\rm AF}\sum_{\langle ij\rangle}{\bm s}_i\cdot{\bm s}_j
-J_{\rm H}\sum_i{\bm S}_i\cdot{\bm s}_i\notag\\
&+g\sum_iQ_iT_i^z+\frac{k}{2}\sum_iQ_i^2+\lambda\sum_{\langle ij\rangle}Q_iQ_j  , 
\label{model}
\end{align}
where the first term represents the Kugel-Khomskii interaction $K$ between the spins ${\bm S}_{i,j}$ 
and the pseudospins ${\bm T}_{i,j}$ in the $d_{xz}$ and $d_{yz}$ orbitals in the nearest-neighbor sites 
${\langle ij\rangle}$; the second and third terms represent the Heisenberg exchange interaction 
$J_{\rm AF}$ between the two spins ${\bm s}_i$ and ${\bm s}_j$ in the $d_{xy}$ orbitals and the Hund's 
ferromagnetic exchange interaction  $J_{\rm H}$, respectively.  
The terms in the last line of Eq.~(\ref{model}) represent the Jahn-Teller coupling between the 
pseudospin and the lattice degrees of freedom written in terms of the normal coordinates of the lattice 
distortion $Q_i$.  

We analyze this model using the semiclassical mean-field approximation \cite{fazekas}, 
taking $\langle S_i^z\rangle$, $\langle T_i^z\rangle$, $\langle s_i^z\rangle$, and $\langle Q_i\rangle$ 
to be the order parameters, where we again assume the extended unit cell of $\sqrt{2}\times\sqrt{2}$.  
The results for the temperature dependence of the order parameters 
are shown in Fig.~\ref{fig5}, where we find that the successive phase transitions are well reproduced, 
i.e., the ordering of the pseudospins occurs at $T_{\rm S}$, which is followed by the ordering of the 
spins at $T_{\rm N}$, just as in the experiment \cite{sakurai}.  The obtained large $T_{\rm S}$ and 
$T_{\rm N}$ are due to the insufficiency of the mean-field approximation, where the strong quantum 
fluctuations of the two-dimensional spin system are completely neglected.  

%%%%%%%%%%%%%%%%%%%
%\section{Summary}
%%%%%%%%%%%%%%%%%%%

In summary, using the DFT-based electronic structure calculations and effective model analysis, 
we have elucidated the origin of the successive phase transitions observed in a layered perovskite 
$\alpha$-Sr$_2$CrO$_4$.  We showed that the crystal-field level of nondegenerate $3d_{xy}$ 
orbitals of the Cr ion is lower in energy than that of doubly degenerate $3d_{yz}$ and $3d_{xz}$ orbitals, 
despite the fact that the CrO$_6$ octahedron is elongated along the $c$-axis of the crystal 
structure, which gives rise to the orbital degrees of freedom in the system with a $3d^2$ electron 
configuration.  This is one of the rare examples of the reversal of crystal-field splitting.  
We obtained the spin- and orbital-ordered ground state by DFT calculations allowing for spin 
polarization.  We proposed the low-energy effective model made of 
the Kugel-Khomskii--type spin-orbital subsystem consisting of the $d_{xz}$ and $d_{yz}$ orbitals of 
the Cr ion and the antiferromagnetic Heisenberg subsystem consisting of the $d_{xy}$ orbitals, 
which are coupled via the on-site Hund's ferromagnetic exchange interaction, together with 
the Jahn-Teller distortion.  
The mean-field calculation of this model was shown to account for the observed successive phase 
transitions.  We showed that the imbalance between the atomic energy levels of the in-plane and 
apical oxygens, as well as the distortion of the CrO$_6$ octahedra, plays an essential role in the 
reversal of the crystal-field levels.  

We hope that our work presented here will encourage further experimental studies, such as the 
direct confirmation of the presence of orbital ordering and the determination of its spatial pattern, 
as well as the precise determination of the low-temperature crystal structure.  

\bigskip
\begin{acknowledgment}
We thank M. Itoh and Y. Ueda for enlightening discussions.  
This work was supported in part by Futaba Electronics Memorial 
Foundation and by a Grant-in-Aid for Scientific Research 
(No.~26400349) from JSPS of Japan.  T.T. acknowledges the support 
from the JSPS Research Fellowship for Young Scientists.  
\end{acknowledgment}

\end{document}